\documentclass[reprint,aps,amssymb]{revtex4-2}

\usepackage{graphicx}% Include figure files
\usepackage{dcolumn}% Align table columns on decimal point
\usepackage{bm}% bold math
\usepackage{amsmath}
\usepackage{graphicx}
\usepackage{subcaption}
\usepackage{amsmath}
\usepackage{enumitem}
\usepackage{amsthm}
\usepackage{xcolor}
\usepackage{mathptmx}
\usepackage[page,toc,titletoc,title]{appendix}
\usepackage{comment}
\usepackage[overload]{empheq}
\usepackage{cases}
\usepackage{lipsum}
\usepackage{amsfonts}
\usepackage{hyperref}
\usepackage{float}
\usepackage{atbegshi}
\usepackage{bm}

\usepackage{subcaption}
\newcommand{\dd}{\mathrm{d}}

\newcommand{\xx}{\bm{x}}

\newcommand{\LLL}{\mathcal{L}}
\newcommand{\LFP}{\mathcal{L}_{FP}}
\newcommand{\FF}{\mathbf{F}}

\def\bi{\begin{itemize}}
\def\ei{\end{itemize}}
\definecolor{rred}{rgb}{0.7,0,0.1}
\definecolor{ccyan}{rgb}{0,.5,1}
\definecolor{greenrb}{rgb}{0.2,0.6,0.2}
\usepackage{hyperref}% add hypertext capabilities
\usepackage[mathlines]{lineno}% Enable numbering of text and display math
\begin{document}

\preprint{APS/123-QED}

\title{The Optimal Growth Mode in the Relaxation to Statistical Equilibrium}
\author{Manuel Santos Guti\'errez}\email{manuel.santos-gutierrez@weizmann.ac.il}
 \affiliation{Department of Earth and Planetary Sciences, Weizmann Institute of Science, Rehovot 76100, Israel}
 
\author{Micka\"el D. Chekroun}%
\affiliation{Department of Earth and Planetary Sciences, Weizmann Institute of Science, Rehovot 76100, Israel}%
\affiliation{Department of Atmospheric and Oceanic Sciences, University of California, Los Angeles, CA 90095-1565, USA}

\date{\today}


\begin{abstract}
Systems far from equilibrium approach stability slowly due to ``anti-mixing" characterized by regions of the phase-space that remain disconnected after prolonged action of the flow.
We introduce the Optimal Growth Mode (OGM) to capture this slow initial relaxation. The OGM is calculated from Markov matrices approximating the action of the Fokker-Planck operator onto the phase space. It is obtained as the mode having the largest growth in energy before decay.  
Important nuances between the OGM and the more traditional slowest decaying mode are detailed in the case of the Lorenz 63 model. The implications for understanding how complex systems respond to external forces, are discussed.
\end{abstract}

\maketitle

In many chaotic and complex systems, mixing plays a crucial role. It describes how the system loses memory of its initial state and reaches a well-defined, statistical steady state. Correlation functions decay then over time during mixing. This decay rate determines two key properties: 1) how quickly the system reaches a statistical steady state \cite{lasota}, and 2) its resilience against external disturbances \cite{lasota,ruelle2009}.

For systems at equilibrium, the approach to a statistical steady state is remarkably fast (exponential), establishing a direct link between mixing, correlations, and the system's response to external forces \cite{kubo,kubo1966}. However, this connection is not guaranteed for systems out of equilibrium. This is because these systems lack smooth invariant probability measures \cite{young2002srb,ruelle_nonequilibrium_1998}. Additionally, the introduction of non-conservative forces violates Liouville's theorem, a cornerstone  of statistical physics \cite[\S I]{landau2013statistical}. As a result, the governing equation (Liouville equation \cite{landau2013statistical}) may lose a key property (self-adjointness), severing the link between the system's long-term and short-term behavior due to the presence of non-normal modes. This phenomenon as been well documented in the fluid dynamics literature \cite{reddy_henningson_1993,farrell_1995,trefethen2005} but its consequences on the action of Liouville propagators---also known as transfer  or Frobenius-Perron operators \cite{schutte1999direct,klus2018data}---in the relaxation to statistical equilibria, has been much less discussed. This paper proposes to bridge this gap.

By hinging on the Ulam method \cite{ulam1960collection}, Markov matrices are instrumental in the learning of transfer operator's properties; see e.g.~\cite{schutte1999direct,junge2009discretization,Froyland2021}.
Markov matrices are also instrumental in analyzing mixing  \cite{meyn2012markov,hairer2010Harris}.
It is thus not surprising that they are used for estimating the long-term behavior of the system's Liouville propagator characterizing how probability densities evolve and revealing how different aspects of the system,
 including correlations, change over time; see e.g.~\cite{froyland2003detecting,Chek_al14_RP, generatorfroyland}. This approach can be justified for both deterministic and stochastic systems  \cite{froylandapproximating1998,fishman2002,Chekroun_al_RP2}. 

Essentially, Markov-Ulam matrices aim at capturing the system's transition probabilities by identifying key regions in its phase space (possible states) and the specific modes (behaviors) that are slowest to lose memory of their initial state  \cite{froyland2005,froyland2009}. They also pinpoint areas most susceptible to external disturbances \cite{froyland2016optimal,Antown2018,froyland2020computation}. Within a Lagrangian setting, these matrices can even identify persistent patterns and barriers to movement within the system's physical space itself  \cite{Froyland2007,froyland2013analytic}.

Despite their relevance to study nonlinear dynamics, the non-uniform exponential decay to equilibrium of Markov-Ulam matrices has been poorly studied. For systems out of equilibrium, these matrices, reflecting the Liouville propagator \cite{schutte1999direct,Chek_al14_RP}, are 
expected to be also non-selfadjoint. As a consequence, non-normal modes are expected to occur: modes that are slow to changes over finite time-intervals and which might not correspond to the slowest (normal) modes to decay in the long run. This paper aims to characterize these modes by introducing a new concept: the Optimal Growth Mode (OGM). In simpler terms, this mode is the one that takes the longest to lose memory and reach equilibrium. As shown below, the OGM is characterized by some sort of ``anti-mixing behavior."

To avoid technical difficulties, we consider deterministic systems embedded within a noise bath. This way, the existence of a unique (and smooth) ergodic statistical steady state can be guaranteed for a broad class of systems \cite{da1996ergodicity,Chekroun_al_RP2}. Thus, our dynamical systems, acting on the $d$-dimensional phase-space $\mathbb{R}^d$, are described by a vector-field $\FF : \mathbb{R}^d \longrightarrow \mathbb{R}^d$ and the following stochastic differential equation (SDE):
 \begin{equation}\label{eq:sto ode 2}
\dot{\xx}  = \FF (\xx) + \sqrt{2 \epsilon} \dot{\bm{W}}_t,
\end{equation}
where $ \dot{\bm{W}}_t$ is an $d$-dimensional white-noise process and $\varepsilon$ is a measure of  its covariance.  
The evolution of probability density functions associated with the SDE \eqref{eq:sto ode 2}, is given by the Fokker-Planck equation (FKE) \cite{risken}:
\begin{equation}\label{eq:fpe 2}
\partial _t\rho = \LFP \rho= -\nabla \cdot \left( \FF\rho \right) + \epsilon \Delta \rho.
\end{equation}
Its solution operator is given by the Fokker-Planck propagator $e^{t\LFP}$ assumed to form a $C_0$-semigroup \cite{engel2000}. When $\epsilon=0$, it reduces to the Liouville propagrator \cite{Chek_al14_RP}.

Under generic conditions on $ \FF $, Eq.~\eqref{eq:fpe 2} has an ergodic statistical steady state $\rho_s$ satisfying $\LFP\rho_s = 0$ and the Fokker-Planck 
 operator $\LFP$ has an isolated spectrum constituted of damped eigenvalues separated from the eigenvalue $0$ (spectral gap); see e.g.~\cite{Chekroun_al_RP2}.  In this case, the underlying damped eigenvalues of the Fokker-Planck 
 operator $\LFP$ are referred to as the Ruelle-Pollicot (RP) resonances and imply an asymptotic exponential decay of correlations \cite{Chekroun_al_RP2}. These resonances hold significant importance. They allow for decomposition formulas for both temporal correlation functions and power spectral density functions \cite[Sec.~2.3]{Chekroun_al_RP2} across a wide range of stochastic systems encountered in physics; see also \cite{ruelle1986resonances,baladi2000positive} for deterministic systems. 
  The spectral gap ensures also the exponential convergence of  $e^{t\LFP}\rho_0$  towards $\rho_s$ (in some $L^2$ space) for a broad class of initial distribution $\rho_0$; see e.g.~\cite[Theorem 6]{Chekroun_al_RP2}. Exponential rate of convergence and spectral gaps also hold  in Wasserstein spaces for an even broader class of models than Eq.~\eqref{eq:sto ode 2}; see \cite{hairer2002exponential,hairer2008spectral}. 
 However the study of transient behavior before reaching exponential convergence remains largely unexplored from an ergodic viewpoint.

In classical fluid dynamics the role of transient dynamics has been thoroughly examined. 
Certain flows experience a burst of transient energy growth before settling into an exponential decay towards steady state, whose rate is controlled by the damped eigenvalues of the linearized governing equations. This temporary spike  is not bounded by these eigenvalues and is caused instead by non-normal modes of the underlying linearized operator \cite{reddy_henningson_1993,farrell_1995,MacTaggart2017,trefethen2005}. Non-normal modes, when influenced by random fluctuations have shown to be highly relevant. They can explain for instance the variability of large-scale climate phenomena like the
El Ni\~no-Southern Oscillation \cite{penland_1993,penland1995optimal}, ocean circulation patterns \cite{tziperman2002transient,zanna2005nonnormal}, or atmospheric flows \cite{farrell1993stochastic,bousquet2015numerical}.

This paper focuses on non-normal modes associated with Fokker-Planck operators $\LFP$, for which exponential relaxation towards a (unique) ergodic statistical equilibrium holds. We are particularly interested to clarify how these modes affect this relaxation process at early times. 

An operator, $\LLL$, is normal if it commutes with its adjoint, $\LLL^{\ast}$ \cite[\S I.6]{kato1966}, namely if  $\LLL^\ast \LLL = \LLL \LLL^\ast$.
Note that self-adjoint operators are a particular class of normal operators. From Liouville's theorem for Hamiltonian systems, the operator $\LFP$ is normal and more generally, for stochastic dynamics within a potential $V$, such that $\mathbf{F}(\xx) = - \nabla V(\xx)$ in Eq.~\eqref{eq:sto ode 2}, it is also true that the Fokker-Planck operator $\LFP$ is normal \cite{pavliotisbook2014}. Operators for which $\LLL^\ast \LLL \neq \LLL \LLL^\ast$  are called non-normal, for which an extensive literature exists \cite{trefethen2005}  albeit not specifically dedicated to the case of Fokker-Planck  operators.

Analytical formulas of RP resonances and modes of the Fokker-Planck operator is restricted to very few instances; see e.g.~\cite{metafunes2002,gaspard1995,TantetJSPII}. In practice, such resonances are numerically estimated from (long) time series $\{ \xx_k\}_{k=1}^T$ obtained by solving Eq.~\eqref{eq:sto ode 2} \cite{crommelin2011diffusion,Chekroun_al_RP2}. 
 As mentioned above, Ulam's method grants a means to obtain estimates of the dynamics' propagator $e^{t\LFP}$  via Markov matrices. The eigenvalues of $\LFP$ are then obtained through logarithm formulas \cite{crommelin2011diffusion,Chekroun_al_RP2}. 
This method consists of subdividing
the phase space $\mathcal{X}$ (typically shadowing the dynamics' attractor)
 into $N$ non-intersecting \emph{cells} or \emph{boxes} $\{B_i\}_{i=1}^{N}$ and estimating the dynamics' probability transitions across these boxes. Mathematically, the propagator $e^{\tau\LFP}$ for a given transition time $\tau$, is approximated by   a $N\times N$ Markov matrix $M_{\tau}$,  whose entries are given by \cite{Chekroun_al_RP2}:
\begin{equation}\label{projected transfer operator}
\left[M_{\tau}\right]_{i,j}=\frac{1}{\rho_0(B_i)}\int_{B_i}e^{\tau \LFP}\chi_{B_j}(\xx)\rho_s(\dd \xx),
\end{equation}
for $i,j=1,\ldots,N$, where $\chi_{B_i}$ denotes the indicator function of the box $B_i$, $\rho_s$ the stationary statistical equilibrium, and $\rho_0$ the initial distribution. The transition matrix is then estimated  by computing a classical maximum likelihood estimator given by \cite{crommelin2011diffusion,schutte1999direct}:
%%%%%%%%%%%%%%%%%%%%%%%%%%%%%%%%%%%%%%%%%%%%
\begin{equation}\label{eq:transition_matrix}
 \left[M_{\tau}\right]_{i,j} = \frac{\#\bigg\{ \Big(\xx_{k}\in B_j \Big) \land \Big(\xx_{k+\ell} \in B_i \Big) \bigg\}}{\# \Big\{\xx_{k} \in B_j\Big\}},
 \end{equation}
 for $i,j = 1,\ldots, N$. 
Given a sampling time $\delta t$ at  which the time series $\xx_k$ is collected, the formula \eqref{eq:transition_matrix} counts the number of observations ($\#\{\cdot\}$)  visiting the box $B_i$ after $\ell= \lfloor \tau/\delta t \rfloor$ iterations, knowing that it already landed in $B_j$.
The resulting operation results into a coarse-graining of the dynamics encoded by $M_\tau$ \cite{Chek_al14_RP,Chekroun_al_RP2} that incorporates artificial diffusion \cite{generatorfroyland,Tantet2018}. Note that for high-dimensional systems, time series constructed from reduced coordinates (observables) are often used instead of the full-state vector $\xx_k$. In this case, transition matrices can still be estimated in a reduced state space and their eigenvalues  are still informative about the genuine system's transitions (in the full state space) through the notion of reduced RP resonances; see \cite[Theorem A]{Chek_al14_RP} and  \cite{Chekroun_al_RP2,RP_ENSO}.

Either estimated in the full or a reduced state space, the transition matrix $M_\tau$ satisfies the Perron-Frobenius theorem \cite{meyn2012markov}. As such, its eigenvalues are contained in the unit ball and there exists an eigenvector $\mathbf{u}$ in  $\mathbb{R}^N$ such that $M_\tau \mathbf{u} = \mathbf{u}$. This vector approximates the ergodic invariant measure of the system.  
 For many stochastic systems, the resulting stochastic Matrix $M_\tau$ is irreducible \cite{Chekroun_al_RP2}, ensuring its dominant eigenvalue $\mu_0$ equal to $1$ to be simple, and the following limit to hold:
\begin{equation}\label{eq:conv_projector}
    \lim _{k \rightarrow \infty}\left\| (M_\tau - E_1)^k  \right\|_2 =0,
\end{equation}
where the matrix $E_1$ is the spectral projector  associated with the eigenvalue $\mu_0$ given by $E_1 = \mathbf{u}\mathbf{1}^{\mathrm{tr}}$, where $\mathbf{1}$ is the $N$-dimensional  (column) vector whose entries are equal to $1$.

Then, the  asymptotic rate of convergence to equilibrium is  determined by the spectral gap $\gamma$ defined as the distance between the subdominant eigenvalue $\mu_1$ and the unit circle, namely $\gamma=1- |\mu_1|$, see \cite{meyn2012markov}. In particular, $\gamma$ controls the  geometric rate of convergence of the matrix $M_\tau$ towards the spectral projector $E_1$ in Eq.~\eqref{eq:conv_projector}.

Such a geometric convergence towards zero of $\delta(k)=\|(M_\tau - E_1)^k  \|_2$ is often characteristic of exponential decay of correlations for the original system; see e.g.~\cite[Sec.~4]{Chekroun_al_RP2}.
 This picture is nevertheless tempered by the potential non-normality of the matrix $M_\tau-E_1$.
In such a case, the relaxation towards zero of $\delta(k)$ takes place only after a few iterations, with some possible transient growth. During this transient, the system  is slow to lose memory and does not exponentially decorrelate in time.

It remains to identify what regions  of  the phase-space---as approximated by the dynamics' coarse-graining tied to Eq.~\eqref{eq:transition_matrix}---display such a transient growth. The resulting mode capturing such regions is expected to indicate sets of initial conditions which remain disconnected for a prolonged interval of time before mixing takes over.

To calculate such a mode, and  inspired by the aforementioned work of \cite{farrell_1995} in fluid dynamics, we seek the directions of maximum stretching upon the action of the relaxation matrix $Q_k=(M_\tau - E_1)^k $. In that respect,  we consider its singular value decomposition (SVD):
\begin{equation}\label{eq:svd}
 (M_\tau - E_1)^k = U_k \Sigma_k V_k^\ast,
\end{equation}
where $U_k$ and $V_k$ are unitary $N \times N$ matrices and $\Sigma_k$ is an $N\times N$ diagonal matrix containing the singular values $\{ \sigma_{k,i} \}_{i=1}^N$, ordered as: $\sigma_{k,1}\geq \sigma_{k,2} \geq \ldots \geq \sigma_{k,N}$. Thus, $\| Q_k  \|_2 = \sigma_{k,1}$ and there exists a vector $\mathbf{v}_{k,1}$ in $\mathbb{R}^N$ such that:
\begin{equation}\label{eq:norm_sing}
    \| Q_k  \|_2=\| Q_k \mathbf{v}_{k,1} \|_2 = \sigma_{k,1}.
\end{equation}
We define then the Optimal Growth Mode (OGM) as the mode that maximizes $\| Q_k  \|_2$ over $k$. This mode, denoted by $\mathbf{w}_{OGM}$ below, is actually given by the vector $\mathbf{v}_{k_0,1}$ that satisfies:
\begin{equation}\label{eq:max_sing}
    \max_{k\geq 0}\| Q_k \|_2 =\|  (M_\tau - E_1)^{k_0} \mathbf{v}_{k_0,1} \|_2 = \sigma_{k_0,1}.
\end{equation}
In other words the OGM maximizes the relaxation curve
\begin{equation} \label{eq:relaxation_map}
k \mapsto \| Q_k \|_2. 
\end{equation}

The OGM must not be confused though with the slowest decaying mode (SDM), defined as the eigenvector $\mathbf{w}_{SDM}$ of $\mathbb{R}^N$ that satisfies:
\begin{equation}\label{eq:sdm}
    M_\tau \mathbf{w}_{SDM} = \mu_1 \mathbf{w}_{SDM}.
\end{equation}
Since  $\lim_{k\rightarrow \infty} \| M_\tau^k \mathbf{w}_{SDM} \|_2=\lim_{k\rightarrow \infty} \mu_1^k = 0 $ and 
$\mu_1$ is the subdominant eigenvalue of $M_\tau$, the SDM is the eigenmode of $M_\tau$ that has the slowest decay at asymptotic times. It indicates regions of phase-space of slow asymptotic mixing. It remains to illustrate how these two modes, the OGM and the SDM, reflect different dynamical properties of chaotic systems.

\emph{The Lorenz 63 system.---}The Lorenz 63 (L63) model  \cite{lorenzdeterministic1963} is a low-dimensional truncation of the equations of atmospheric convection \cite{ghil2012topics}, whose dynamical properties have been extensively studied \cite{sparrow}. 
In its  standard chaotic regime, statistical properties have been also analyzed \cite{anishchenko2005statistical,Chekroun2011}.  Linear response has been shown  to hold (on numerical grounds), i.e.~small changes in parameter forcing equate to proportional changes in the statistical moments \cite{reicklinear2002, Lucarini2016, SantosJSP}. The L63 system writes as follows:
\begin{equation}\label{lorenz63}
\dot{\mathbf{x}}(t)=\mathbf{F}(\mathbf{x})=\begin{cases} s (y - x) \\ x(r - z) - y  \\ xy -b z  \end{cases},
\end{equation}
where the model parameters are set to be $s=10$, $r=28$ and $b=8/3$, for which the system displays its famous ``butterfly strange attractor" \cite{Tucker2002}.

To estimate the Markov matrix $M_\tau$ associated with
Eq.~\eqref{lorenz63}, we integrate it over $10^5$ time units (after spinup) with a time step of $\delta t = 10^{-3}$ time units, using a fourth order Runge-Kutta algorithm. Thus, a total of $T = 10^8$ data points in phase-space is recorded. The trajectory $\{(x_k,y_k,z_k)\}_{k=0}^{T}$ shadows an attractor contained in a compact subset of $\mathbb{R}^3$ \cite{sparrow}, which lies in the domain $[-20,20]\times [-30,30]\times [0,50]$.  We work then within the reduced state consisting of the $(x,z)$-domain, $[-20,20]\times [0,50]$.
This domain is divided into $N = 2^{12}$ equally sized, non-intersecting boxes $\{B_i\}_{i=1}^N$ of dimensions $40/2^6 \times 50/2^6$. Then, the transition time $\tau = 10\cdot \delta t$ is selected and the transition matrix $M_\tau$ given in Eq.~\eqref{eq:transition_matrix} is computed with $\bm{x}_k=(x_k,z_k)$. While the L63 system is deterministic, the coarse box-resolution yields artificial diffusion \cite{Tantet2018} analogous to the addition of small noise to Eq.~\eqref{lorenz63}.

In the reduced state space focusing only on variables $x$ and $z$,
 the eigenvalues of the estimated matrix $M_\tau$ correspond to simplified versions of resonances of the full transfer operator (reduced RP resonances  \cite{Chekroun_al_RP2}). These reduced resonances are linked to the  conditional expectation \cite{Chekroun_al_RP2} in the equation closure of Eq.~\eqref{lorenz63} within this reduced state space \cite{kondrashovdata2015,LC2023}. This conditional expectation represents the behavior of the system after averaging out unobserved variables. However, it may not capture all important aspects of the system's variability \cite{chekroun2016emergence}.  As shown in the appendix, the spectrum of the reduced Markov matrix $M_\tau$ allows us here to approximate the L63 model's temporal variability, as reflected by its correlation functions, with reasonable accuracy. This suggests that the modes obtained from the reduced transition matrix $M_\tau$ capture, in the reduced state space, actual dynamical properties of the L63 model. This is confirmed as explained below. 

%%%%%%%%%%%%%%%%%%%%%%%%%%%%%%%%%%%%%%%%%%
\begin{figure}
\centering
    \includegraphics[height=0.18\textwidth, width=.5\textwidth]{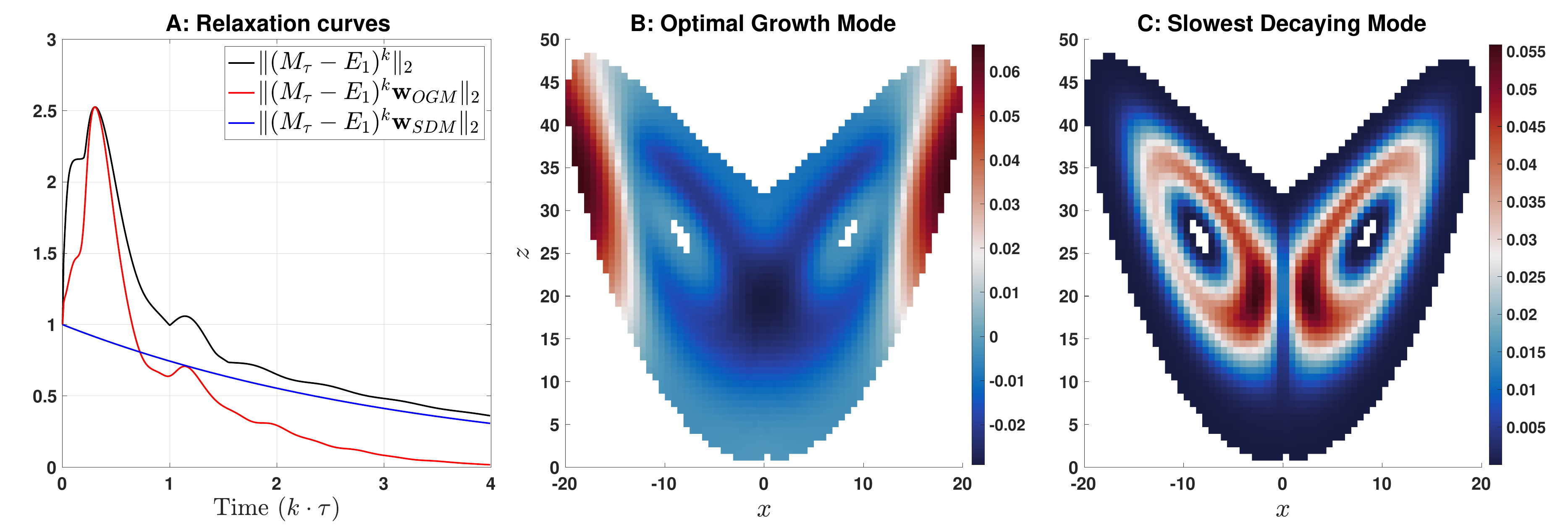}
\caption{\label{fig:1}\textbf{Panel (A):} the black curve shows the relaxation curve \eqref{eq:relaxation_map}. The red and blue curves are the successive evaluation of the matrix $M_\tau - E_1$ applied on the OGM and SDM, respectively. \textbf{Panel (B):} OGM calculated using the SVD of the reduced transition matrix $M_\tau$. \textbf{Panel (C):} SDM (modulus) calculated from the eigen-decomposition of the reduced transition matrix  matrix $M_\tau$.}
\vspace{-2ex}
\end{figure}
%%%%%%%%%%%%%%%%%%%%%%%%%%%%%%%%%%%

%%%%%%%%%%%%%%%%%%%%%%%%%%%%%%%%%%%%%%%%%%
\begin{figure*}
\centering
    \includegraphics[height=0.6\textwidth, width=1\textwidth]{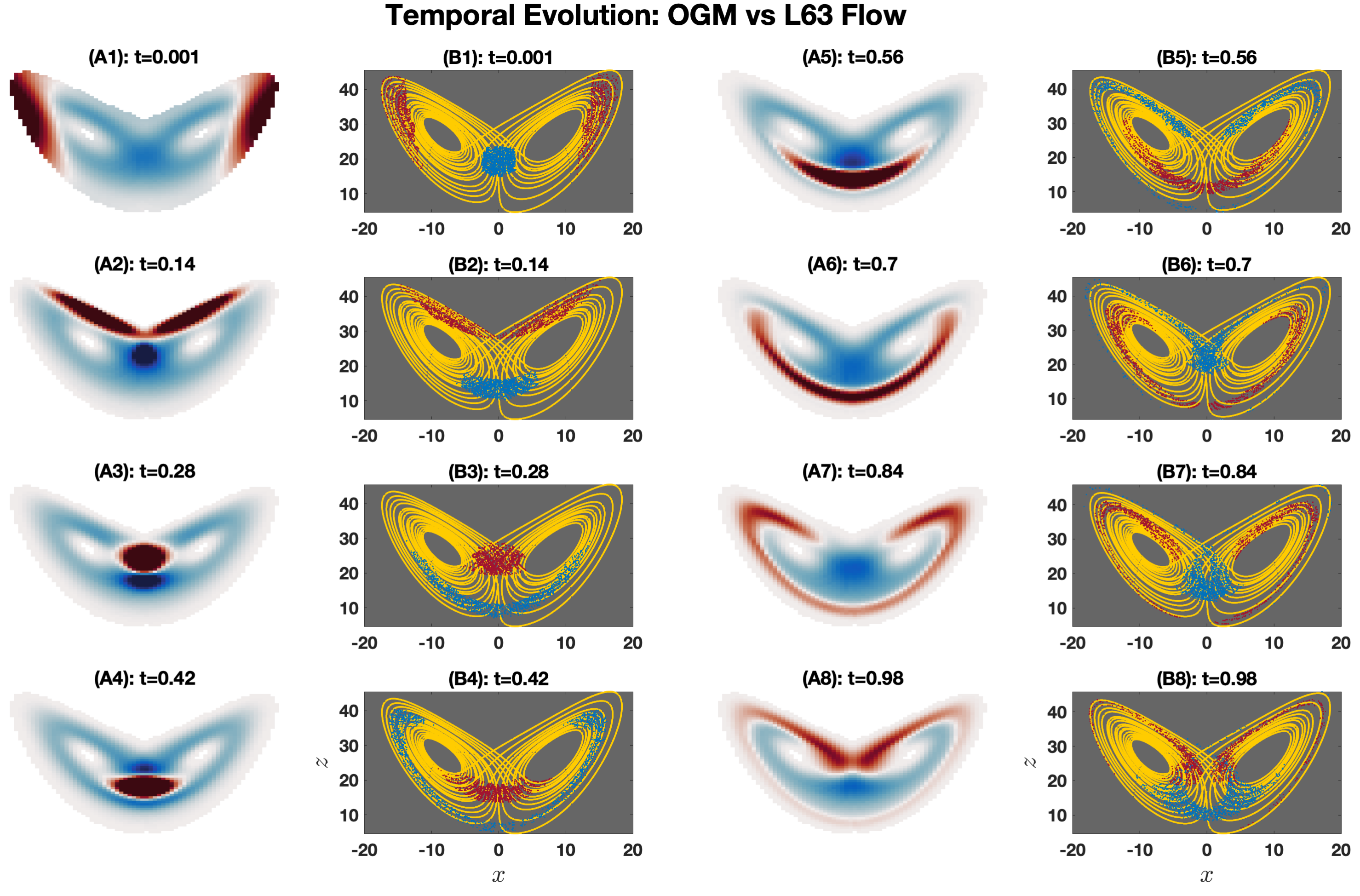}
\caption{\label{fig:2} \textbf{Evolution of the OGM.} The A-panels show $(M_\tau-E_1)^k\mathbf{w}_{OGM}$, for $k=0,14,\ldots,98$. The B-panels show the evolution of initial conditions seeded around the positive and negative areas of the OGM and evolved in time for the intervals indicated in the titles.}
\vspace{-1ex}
\end{figure*}
%%%%%%%%%%%%%%%%%%%%%%%%%%%%%%%%%%%%%%%%%%%%%
The relaxation curve (Eq.~\eqref{eq:relaxation_map}) is shown as the solid black line in Fig.~\ref{fig:1} and clearly shows the signature of non-normal transient growth and sub-exponential decay, reaching a peak at $t\approx 0.3$ time units. Only for larger times the leading eigenvalue $\mu_1$  takes over, allowing for the expected exponential rate of decay  to take place. The OGM (Fig.~\ref{fig:1}B) follows an evolution in time (red curve in Fig.~\ref{fig:1}A) that reaches its maximum at around $0.3$ time units while the SDM (Fig.~\ref{fig:1}C) does not show transient growth (blue curve in Fig.~\ref{fig:1}A) although it corresponds to  the slowest decaying mode at large times.

Figure \ref{fig:2} shows the evolution of $\left(M_\tau - E_1 \right)^k\mathbf{w}_{OGM}$ over time.
The panels (A) depict this for different time instants (see subtitles) corresponding to different values of $k$ $(0, 14,\ldots, 98)$. Panels (B) show the fate of $10^{3}$ red and $10^{3}$ blue trajectories over time. The red trajectories start in areas where normalized OGM is close to $0.05$, and the blue trajectories start in areas where it is close to $-0.05$.
We can see that the OGM, which initially concentrates near the top of the attractor, moves towards the center and collapses before expanding outwards again. Interestingly, the first signs of this expansion occur between $t= 0.3$ and $t=0.35$, which coincides with the peak of the relaxation curve. The trajectories (red and blue patches in panel B) also contract and expand with the attractor, but importantly, they do not mix with each other until $t=0.98$. This suggests that the OGM can effectively identify regions of the attractor that mix slowly.

\begin{figure*}
\centering
    \includegraphics[height=0.6\textwidth, width=1\textwidth]{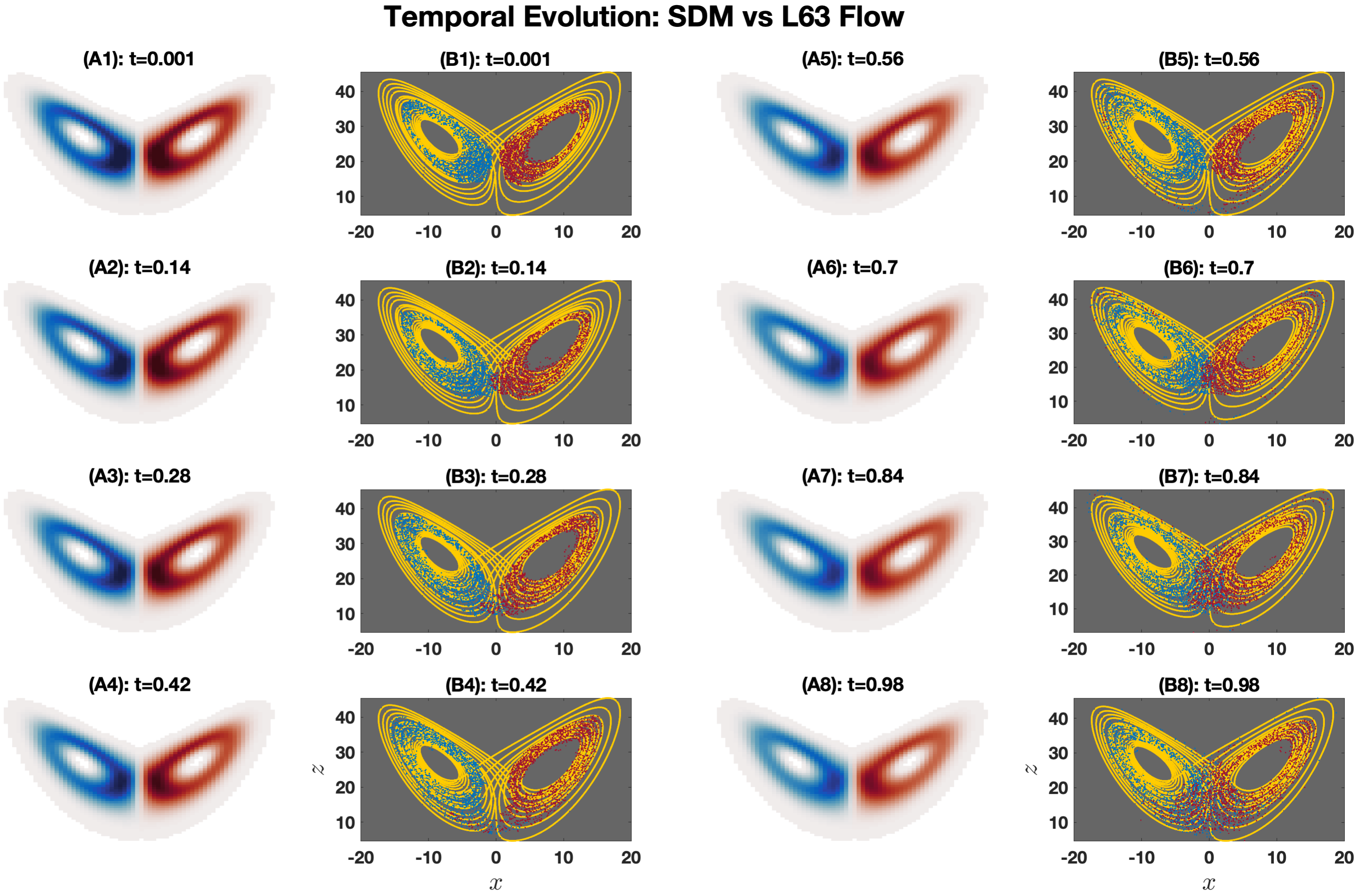}
\caption{\label{fig:3} \textbf{Evolution of the SDM.} In the A-panels we show $(M_\tau-E_1)^k\mathbf{w}_{SDM}$, for $k=0,14,\ldots,98$. The B-panels show the evolution of initial conditions seeded around the positive and negative areas of the SDM and evolved in time for the intervals indicated in the titles.}
\vspace{-1ex}
\end{figure*}

In contrast to OGM, the SDM (whose absolute value is shown in Fig.~\ref{fig:3}) appears to remain unchanged over time. 
Its shape is reminiscent to that of the almost invariant set calculated in \cite{froyland2009almost} for the full 3D model.
This is evident not only in panels A of Figure \ref{fig:3}, but also in the trajectories (red and blue patches in panel B).
 These trajectories mostly stay within their respective positive and negative SDM patches. Additionally, unlike the OGM, the trajectories colored blue  and red mix much faster. By $t = 0.14$, we observe for instance particles of both colors crossing the $x=0$ line in both directions.

Both OGM and SDM exhibit properties that slow down mixing in the system, but in different ways.
Initially, the OGM experience energy growth at early times. However, as time progresses, its convergence to a stable, long-term state (ergodic steady state) might become faster. The SDM experiences a steady decrease over time, following an exponential function with a rate determined by a specific time scale given as $\tau^{-1}\log(|\mu_1|)$ \cite[Eq.~(3.40)]{Chekroun_al_RP2}. However, by design, the SDM is guaranteed to decay the slowest compared to other modes at very long times. This behavior is evident in Figure \ref{fig:1}A for the L63 system. Furthermore, trajectories starting in regions with opposite OGM values mix much slower than those starting in opposite SDM regions. This is illustrated by comparing the colored patches in Figures \ref{fig:2}B and \ref{fig:3}B.

%%%%%%%%%%%%%%%%%%%%%%%%%%%%%%%%%%%%%%
%%%%%%%%%%%%%%%%%%%%%%%%%%%%%%%%%%%%%%

\emph{Discussion.}--- 
Systems out of equilibrium, characterized by non-normal equations of motion, exhibit uneven mixing in phase space. This prevents rapid convergence to a statistical steady state initially, even if the system eventually decorrelates exponentially fast.
The SDM (Eq.~\eqref{eq:sdm}) captures the slowest asymptotic decay, but it misses the crucial short-term "anti-mixing" behavior common in non-normal Fokker-Planck operators (Eq.~\eqref{eq:fpe 2}). 
On the contrary, the OGM captures transient dynamics. This work introduces an algorithm based on SVD decomposition of Markov-Ulam matrices to compute the OGM.
 Unlike the SDM, the OGM exhibits growth in its norm at early times, peaking at the maximum point in the relaxation curve (Eq.~\eqref{eq:relaxation_map}). This anti-mixing character of the OGM manifests in two ways: (i) {\it sharp increase}---
the OGM norm rises significantly before decaying, and (ii) {\it slow mixing trajectories}---trajectories starting in regions with opposite OGM values mix much slower compared to, for example, those in the near-invariant regions identified by the SDM (Figures \ref{fig:2}B and \ref{fig:3}B).

Transient slow mixing has consequences for the system's response to perturbations.
 A slow approach to equilibrium translates to increased sensitivity to perturbations due to the slow decay of the Green function \cite{ruelle2009,Chek_al14_RP,lucarini2008,Santos_Gutierrez_2022}. We expect the OGM to be associated with perturbations that take longer to enter the exponentially decaying regime, as suggested by the L63 system's relaxation curve (Fig.~\ref{fig:1} A) and the system's non-equilibrium nature.
The OGM is optimized over time for transient behavior, so it may not be ideal for optimizing the steady-state response to perturbations. Prior work has addressed finding optimal perturbations in the linear response regime (\cite{Antown2018,Schweitzer1968,Lucarini2016}). Our future work will explore the implications of non-normality for linear response in more detail.

%\begin{acknowledgements}
\vspace{2ex}
{\small \emph{Acknowledgements.}---MSG is grateful to the Feinberg Graduate School for their support through the Dean of Faculty Fellowship. 
This work has been supported by the Office of Naval Research (ONR) Multidisciplinary University Research Initiative (MURI) grant N00014-20-1-2023, and by the National Science Foundation grant DMS-2407484.
This work has been also supported by the European Research Council (ERC) under the European Union’s Horizon 2020 research and innovation program (grant agreement No. 810370). 
%\end{acknowledgements}

The authors declare that there are no competing interests.}

\begin{figure}
\centering
    \includegraphics[height=0.35\textwidth, width=.48\textwidth]{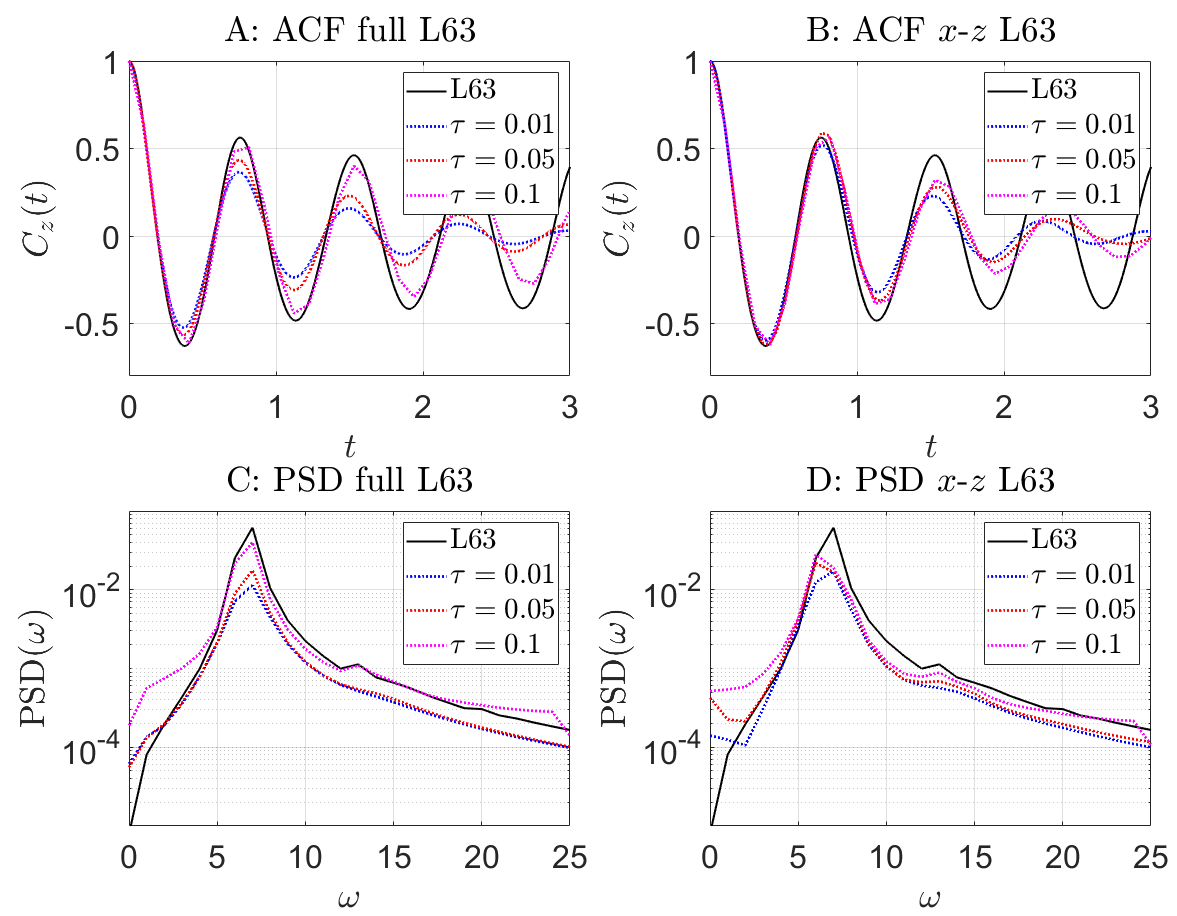}
    \caption{\label{fig:l63_acf_psd} Panels A and C show the autocorrelation function and power spectral density of the variable $z$ for the L63 model. The black lines are obtained from the integration of the full model. The dotted lines correspond to the correlation functions as computed from the spectrum of the (full) transition matrix $M_\tau$ (Eq.~\eqref{eq:transition_matrix} with $\bm{x}_k=(x_k,y_k,z_k)$), for values of $\tau$ as indicated in the legend. Panels B and D, show correlation functions and PSDs as computed from the spectrum of the reduced  transition matrix  (Eq.~\eqref{eq:transition_matrix} with $\bm{x}_k=(x_k,z_k)$), estimated in the $(x,z)$-reduced state space.}
\end{figure}

\vspace{2ex}
%\begin{comment}
{\small 

\emph{Appendix: Correlations approximation.}--- In computing $M_\tau$ in Eq.~\eqref{eq:transition_matrix}, there are two main parameters that affect the computations: the number of boxes $N$, and the transition time $\tau$. While increasing $N$ improves the approximation \cite{fishman2002}, in this appendix, we shall illustrate on the L63 system how $\tau$ influences the approximation of correlation functions and power-spectra using the decomposition formulas of \cite[Sec.~2.3]{Chekroun_al_RP2}. We furthermore compare the performance when the underlying resonances are estimated from 
the (3D) transition matrices $M_\tau$ (Eq.~\eqref{eq:transition_matrix} with $\bm{x}_k=(x_k,y_k,z_k)$) versus those obtained from the reduced transition matrix (Eq.~\eqref{eq:transition_matrix} with $\bm{x}_k=(x_k,z_k)$), estimated in the $(x,z)$-reduced state space.  For the computations in 3D, the $(x,y,z)$-domain is $[-20,20]\times [-30,30] \times [0,50]$ that is divided into $N = 2^{15}$ equally sized, non-intersecting boxes $\{B_i\}_{i=1}^N$ of dimensions $40/2^5\times 60/2^5 \times 50/2^5$. Then, the full transition matrix $M_\tau$ is computed for different values of $\tau$ as detailed below.

In Figure~\ref{fig:l63_acf_psd}A we show the autocorrelation functions (ACFs) of the $z$ variable of the L63 model in a number of ways. The black solid line is calculated as the sample autocorrelation function of a $10^5$ time unit time series, and the dotted lines are computed using Markov matrices over the three-dimensional attractor, with values of $\tau$ indicated in the legend. In Panel B, dotted lines are computed in the  $(x,z)$-reduced state space. In Panels C and D, we show the corresponding power spectral densities (PSDs). The results show that the oscillatory component of the correlation function is well captured regardless of the value of $\tau$; see the peaks in such panels. The readouts also suggest that an increasing value of $\tau$ improves the correlation function approximation, yet it compromises the approximation of the  low-frequency variability portrayed by the PSDs at the smallest $\omega$-values in panels C and D. This surprising ability of the Markov chain method to capture the frequency of oscillations was also reported for stochastic differential equations in \cite[Sec.~4]{Chekroun_al_RP2}. 

}

%\end{comment}

% \nocite{*}

\bibliography{apssamp_MC}% Produces the bibliography via BibTeX.

\end{document}